\documentclass[journal]{IEEEtran}
\ifCLASSINFOpdf
\else
\fi
\usepackage{graphicx}
\usepackage{amssymb}
\usepackage{cite}
\usepackage{amsmath,amsfonts,algorithmic}
\begin{document}
%
\title{Unsupervised cross-user adaptation in taste sensation recognition based on surface electromyography with conformal prediction and domain regularized component analysis}
%
%
%

\author{Hengyang Wang, Xianghao Zhan, Li Liu, Asif Ullah, Huiyan Li, Han Gao, You Wang, Ruifen Hu*, Guang Li
\thanks{H. Wang, X. Zhan, L. Liu, A. Ullah, H. Li, You. W, R. Hu, and G. Li are with the State Key Laboratory of Industrial Control Technology, Institute of Cyber-Systems and Control, Zhejiang University, Hangzhou 310027, China.(Corresponding Author: R. Hu e-mail: 0011377@zju.edu.cn)}
\thanks{This research was supported by the Science Foundation of Chinese Aerospace Industry under Grant JCKY2018204B053 and the Autonomous Research Project of the State Key Laboratory of Industrial Control Technology, China (Grant No. ICT2021A13). }
\thanks{H. Wang, X. Zhan and L. Liu contributed equally to this work.}}

%
%

\markboth{IEEE Transactions on Instrumentation and Measurement}%
{Shell \MakeLowercase{\textit{et al.}}: Unsupervised cross-user adaptation in taste sensation recognition based on surface electromyography with conformal prediction and domain regularized component analysis}
%



\maketitle

\begin{abstract}
Human taste sensation can be qualitatively described with surface electromyography. However, the pattern recognition models trained on one subject (the source domain) do not generalize well on other subjects (the target domain). To improve the generalizability and transferability of taste sensation models developed with sEMG data, two methods were innovatively applied in this study: domain regularized component analysis (DRCA) and conformal prediction with shrunken centroids (CPSC). The effectiveness of these two methods was investigated independently in an unlabeled data augmentation process with the unlabeled data from the target domain, and the same cross-user adaptation pipeline were conducted on six subjects. The results show that DRCA improved the classification accuracy on six subjects (p $\textless$ 0.05), compared with the baseline models trained only with the source domain data, while CPSC did not guarantee the accuracy improvement. Furthermore, the combination of DRCA and CPSC presented statistically significant improvement (p $\textless$ 0.05) in classification accuracy on six subjects. The proposed strategy combining DRCA and CPSC showed its effectiveness in addressing the cross-user data distribution drift in sEMG-based taste sensation recognition application. It also shows the potential in more cross-user adaptation applications. 
\end{abstract}


%
\IEEEpeerreviewmaketitle

\section{Introduction}
%
%
%
%
\IEEEPARstart{T}{aste} sensation is one key component of human feedback, which is not only closely related to the perception and evaluation of food quality and safety \cite{Wang1}, but also related to an individual's health status. For example, cancer patients, diabetes patients and COVID-19 patients might have disordered taste sensations as one of the complication symptom. \cite{taste_clinical,taste_clinical_2,taste_clinical_3}. To objectively and quantitatively describe human taste sensation, recently, researchers have developed systems based on surface electromyography (sEMG) and machine learning models to recognize human taste sensations \cite{Wang, Wang32}. This approach enables the researchers to recognize the taste sensation without verbal expression from the subjects, which can be very helpful in objectively detecting and then improving the feeding experience of those who lost their verbal abilities due to accidents, heredity and diseases, such as those suffering traumatic brain injury \cite{zhan2021rapid} in the ICU who cannot verbally express their taste feelings. For example, our previous research have reached a five-fold cross-validation accuracy of 74.46\% on classifying six types of human basic taste sensations \cite{Wang}.

Although previous studies have shown the effectiveness of pattern recognition of human taste sensations with sEMG, the studies only show high accuracy of taste sensation recognition on the test data from the same subjects from whom the training data were collected \cite{Wang} even though cross-validation was used. Although this ideal pattern of sEMG-based taste sensation recognition application can be realized in the laboratory environment where all the users have their own database built and models specific to each user can be developed, the cross-user data drift in the real-world applications may pose a huge challenge to the application of this type of system. When the sEMG-based taste sensation recognition system is applied in the real-world scenarios, the new test data collected by the users may bear statistically significantly different distribution from those collected by volunteers in the laboratory due to the different environmental conditions, different skin physical and electronic properties of each individuals, as well as sensor aging and sensor drifts. Therefore, the cross-user domain drift may lead the models trained on the data collected from specific volunteers in the laboratory environment to fail in the recognition task. For example, in our experiments \cite{Wang}, we have found that the accuracy of the model trained on one volunteer's data significantly deteriorates on the test data on the other users.

Domain drift has been a well-known issue in the instrumentation and measurement technology development \cite{zhang2017anti,domain_transfer_sEMG}. Multiple domain adaptation (DA) technologies have been developed by researchers, such as ensemble classifiers, semi-supervised learning and drift correction \cite{zhang2017anti,21,22,23,24,25,26} but the effect can be limited. Recently, the domain regularized component analysis (DRCA) has been proposed by researchers for drift compensation \cite{zhang2017anti}. This approach applies linear projection to find a lower-dimensional hyper-plane which minimizes the difference between the source domain data and the target domain data. Under the assumption that the distribution from either the source domain or the target domain is relatively consistent, this approach has been shown effective in electronic-nose-based gas recognition tasks \cite{zhang2017anti}. Considering the domain drift in sEMG-based taste recognition is generally caused by the subject changes, we believe the data distribution consistency assumption can be satisfied. Therefore, we investigated whether the cross-user data drift in this task can be compensated by the DRCA so that we can reach a higher accuracy on the test data from another subject.

In addition to DRCA, another approach proves to be effective in significantly boosting machine-learning-based pattern recognition performance under sensor drifts: the reliability-based data augmentation with unlabeled data \cite{liu2021cpsc,liu2021boost}. Starting from the models trained on the labeled training data, this strategy leverages the unlabeled data in the target domain where the test data come from, and applies the conformal prediction (CP) to filter the model predictions based on the prediction reliability to be the training data augmentation. The unlabeled data enable the models to gradually adapt to the test data distribution (which is also the unlabeled data distribution) in an unsupervised manner. We have shown in two of our previous studies that this reliability-based data augmentation strategy enables statistically significant model accuracy improvement on electronic-nose-based herbal medicine classification problems \cite{liu2021cpsc,liu2021boost}, where our novelly proposed conformal prediction with shrunken centroids (CPSC) optimized the process with both significantly higher accuracy and shorter computation time.

In this study, to address the cross-user data drift problem in sEMG-based taste sensation recognition task, both DRCA and CPSC were leveraged on the sEMG data we collected from seven different volunteers. We simulated the cross-user data drift by regarding the data collected on one volunteer as the laboratory-based labeled training data (the source domain), and regarding the data from other users as the unlabeled data (the target domain). Then, we compared the effectiveness of DRCA, CPSC, and DRCA coupled with CPSC against the baseline models trained merely on the labeled training data on the source domain.

\section{Materials and Methods}

\subsection{Experiment and Dataset}
\begin{figure*}

    \centering
    \includegraphics[width=0.5\linewidth]{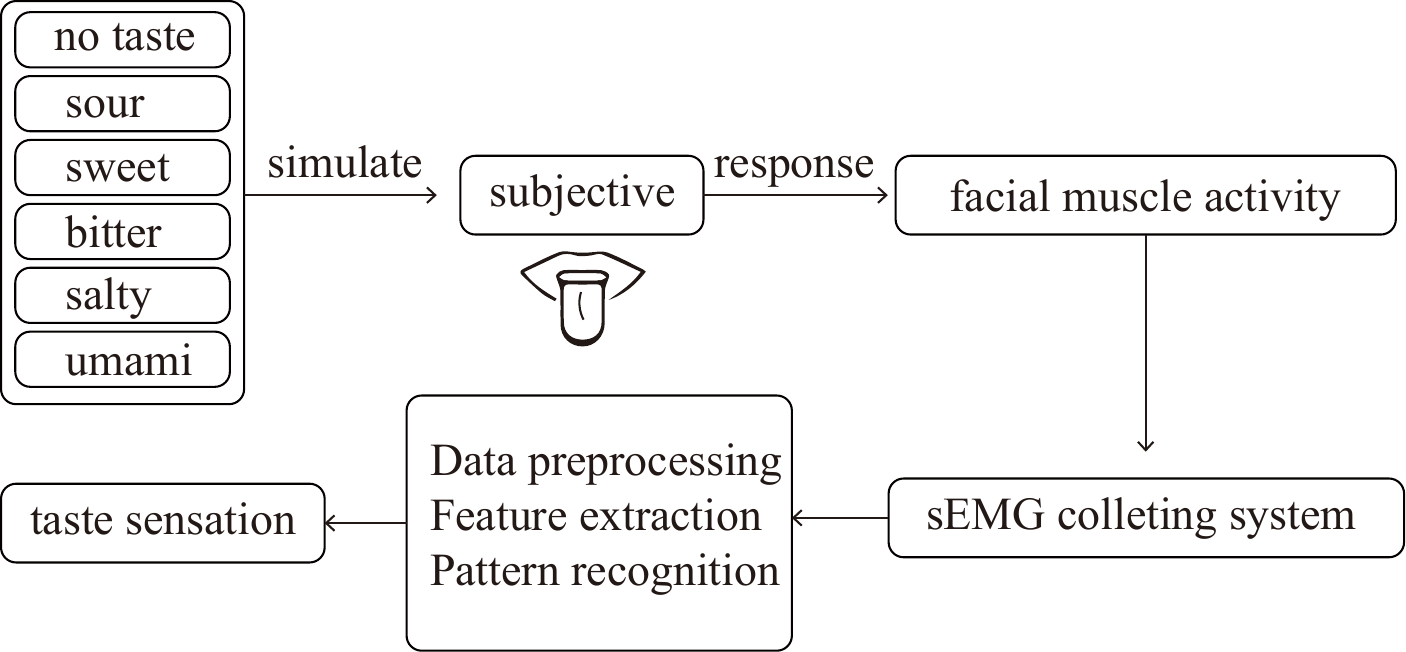}
    \caption{\textbf{The diagram of the taste sensation experiment.} The subjects are provided with different taste stimuli, and an sEMG collecting system is used to record the electrophysiological activities of their facial muscles. After data preprocessing, feature engineering, and pattern recognition, the sEMG signals are discriminated into different taste sensations.}
    \label{experiment diagram}
\end{figure*}

The experiments were designed to classify six basic taste sensations: no taste, sour, sweet, bitter, salty, and umami. These tastes are respectively stimulated by distilled vinegar (Hengshun, with total acid content more than 5 g per 100 mL), white granulated sugar (Taigu, whose sucrose ratio was not less than 99.6\%), instant coffee powder (UCC117), refined salt (Zhongyan, with sodium chloride ratio more than 98.5\%), and Ajinomoto (Xihu, with sodium glutamate ratio more than 99.9\%) \cite{Wang}. Seven volunteers (3 females and 4 males) were involved as the experiment subjects, and their ages range from 22 to 25 years, with an average of 23. This research has been approved by the Ethics of Human and Animal Protection Committee of Zhejiang University. The subjects were in healthy state, as they declared no medication, disease, or any case about taste disorders during the whole period of experiments.\\
\indent For each subject, the cross-user recognition of taste sensation follows two main stages: 1) with every taste stimulation, the sEMG system collects signals with the electrodes on the subject's facial skin; 2) features are extracted from the collected multi-channel signals, and then classified with pattern recognition algorithms (CPSC is adopted in this study, please refer to Section 2.4), as Fig \ref{experiment diagram} shows.\\

\indent To avoid the interference from irrelevant factors which may contribute to noisy sEMG signals, the subjects were asked not to eat anything since 1 hour before each experiment, and they were also suggested not to eat food with strong taste in 24 hours before each experiment. \\
\indent The sEMG data collecting system is made by non-invasive and standard Ag/AgCl electrodes, and they are placed on the parotid glands and muscles (masseters, depressor anguli oris, and depressor labii inferioris), which are associated with salivary secretion and facial expression after taste stimulation\cite{Horio,Wang}. The positions of the electrodes on facial muscle are listed in Table. \ref{electrode positions}, and more details can be found in our previous publication \cite{Wang}. Two electrodes are set to be the bias electrode ('B') and reference electrode ('R'), and the remaining six electrodes collect signals from different channels, with a sampling frequency of 1000 Hz.   \\
\indent In a session of the experiments, six taste stimulation experiments were applied with a random order. The average number of sessions the volunteers took is 65, and each session of a specific taste stimulation consists 11 steps:
\begin{table}[]

\caption{{The positions and types of electrodes on facial muscle}}
\begin{tabular}{cccll}
\cline{1-3}
Electrode Indice     & Type of electrode            & Attached facial muscle                                                          &  &  \\ \cline{1-3}
1                    & Different electrode pairs    & Left masseter                                                             &  &  \\
2                    & Differential electrode pairs & Left masseter                                                             &  &  \\
3                    & Single electrode             & Left depressor anguli oris                                                &  &  \\
4                    & Single electrode             & \begin{tabular}[c]{@{}c@{}}Left depressor\\ labii inferioris\end{tabular} &  &  \\
5                    & Single electrode             & Right masseter                                                            &  &  \\
6                    & Single electrode             & Right masseter                                                            &  &  \\
B                    & Bias electrode               & Left mastoid                                                              &  &  \\
R                    & Reference electrode          & Right mastoid                                                             &  &  \\ \cline{1-3}
\multicolumn{1}{l}{} & \multicolumn{1}{l}{}         & \multicolumn{1}{l}{}                                                      &  &  \\
\multicolumn{1}{l}{} & \multicolumn{1}{l}{}         & \multicolumn{1}{l}{}                                                      &  & 
\end{tabular}
\label{electrode positions}
\end{table}

\begin{enumerate}
    \item The sEMG collecting system was carefully checked by an experiment assistant.
    \item The experiment assistant pasted the electrodes on the volunteer's cleaned face.
    \item The volunteer was asked to stay in a relaxed state and not to constrain his/her facial expression, with their eyes closed and mind calmed down. The experiment assistant started to record the sEMG signals for 12 seconds, whose label belongs to the category 'no taste'.
    \item The volunteer was asked to gargle for 5 times, with 20 mL purified water for each time.
    \item The volunteer was asked to relax for 2 minutes.
    \item The volunteer pushed out his/her tongue, and the experiment assistant placed a level spoon of taste stimulation substance (sugar, coffee powder et. al) on the tip of the tongue. The volume of the spoon was 2 mL.
    \item The volunteer took back his/her tongue, closed the mouth, and kept the tongue staying still. Then he/she stayed in a relaxed state and naturally react with facial expressions, with their eyes closed while focusing solely on the taste sensation. The experiment assistant runs the sEMG collecting system to record the signals for 12 seconds, labelling its taste category.
    \item Repeat step 4 and 5 to clean the tongue of volunteer.
    \item Repeat step 6 with the next taste stimulation.
    \item The volunteer was asked to take a break for 10 minutes.
    \item Repeat step 3-10 for the next session.

\end{enumerate}
\subsection{Feature Extraction}

\subsubsection{Data preprocessing}
\indent Since the facial muscles react continuously with a taste stimulation \cite{Wang}, firstly, the multi-channel signals recorded by sEMG collecting system are partitioned into samples with a sliding window to augment the data. Each sample was cut with a length of 1 second, and the window has a step length of 0.25 second.\\
\indent To attenuate the affect from irrelevant components in the signals, two types of noises are removed: baseline drift and power frequency interference. The baseline drift is due to the zero drift of the device and change in facial muscle tension. We consider the baseline drift as a quadratice polynomial trend term, and it is removed by a fourth-order high-pass Butterworth filter, with a cutoff frequency of 10 Hz. To get rid of the power frequency interference, the samples are processed 
with an adaptive notch filter, which utilizes a high amplitude at 50 Hz and harmonics and remains the most normal signals. After the data preprocessing, several samples with severe distortion were dropped out. The numbers of valid samples from each subject after data preprocessing are listed in Fig. \ref{study pipeline}



\subsubsection{Feature extraction}
The feature extraction process was done on both frequency domain and time domain\cite{EMG_feature_2}. For each channel, 4 frequency features were extracted: the integral values of amplitude, root mean square frequency (RMSF) the frequency centroid (FC), and root var frequency (RVF) \cite{EMG_feature}. As shaped in circles, the electrodes have lowpass characteristic \cite{Wang}. According to the signal spectrum, the energy of sEMG data was indeed denser in the window from 10 Hz to 100 Hz. Therefore, the intervals with variant lengths were used to clip the signals. For the channels in each sample, the spectrum was clipped into 13 intervals to extract the integral values. The interval length was set to be 10 Hz (from 10 Hz to 100 Hz) and 100 Hz (from 100 Hz to 500 Hz) respectively, as Equation. \ref{integral} shows. In the time domain, five features were extracted: root mean square (RMS), zero-crossing rate(ZCR), mean absolute value (MAV), kurtosis (Ku), and skewness (Kw) \cite{EMG_feature_3}. In total, 21 features were calculated for each channel ($F[n]$ denotes the $n-th$ feature), where $F[1]-F[16]$ are frequency domain features and $F[17]-F[21]$ are time domain features:

\begin{equation}
\label{integral}
F[n]=\left\{\begin{array}{l}
\frac{1}{10} \sum_{i=n \times 10}^{(n+1) \times 10-1} f(i), \text { when } 1 \leq n \leq 9 \\
\frac{1}{100} \sum_{i=(n-9) \times 100}^{(n-8) \times 100-1} f(i), \text { when } 10 \leq n \leq 13
\end{array}\right.
\end{equation}

\begin{equation}
    F[14]=\frac{\sum_{i=1}^{L / 2} f(i) \times\left(i \times f_{s} / L\right)}{\sum_{i=1}^{L / 2} f(i)} 
\end{equation}

\begin{equation}
    F[15]=\sqrt{\frac{\sum_{i=1}^{L / 2} f(i) \times\left(i \times f_{s} / L\right)^{2}}{\sum_{i=1}^{L / 2} f(i)}}
\end{equation}

\begin{equation}
    F[16]=\sqrt{\frac{\sum_{i=1}^{L / 2} f(i) \times\left(i \times f_{s} / L-F[14]\right)^{2}}{\sum_{i=1}^{L / 2} f(i)}}
\end{equation}
\begin{equation}
    F[17]=\sqrt{\frac{\sum_{i=1}^{L} x_{i}^{2}}{L}} 
\end{equation}

\begin{equation}
    F[18]=\sum_{i=1}^{L-1} 1, \text { when } x_{i} x_{i+1}<0
\end{equation}

\begin{equation}
    F[19]=\frac{\sum_{i=1}^{L}\left|x_{i}\right|}{L} 
\end{equation}

\begin{equation}
    F[20]=\frac{1}{L} \sum_{i}^{L}\left(\frac{x_{i}-\mu}{\sigma}\right)^{4} 
\end{equation}

\begin{equation}
    F[21]=\frac{1}{L} \sum_{i}^{L}\left(\frac{x_{i}-\mu}{\sigma}\right)^{3}
\end{equation}
In frequency domain, $f(i)$ denotes the spectrum amplitude value of the $i$-th frequency point, and $L$ denotes the length of the channel (1000 in this case), and $f_{s}$ marks the sampling frequency (1000 in this case). In time domain, $x_{i}$ denotes the value of the signal value at $i$-th point. $\mu$ and $\sigma$ are the mean and the standard deviation of the data at sampling points.

\subsection{Study Pipeline}

\begin{figure*}
    \centering
    \includegraphics[width=0.8\linewidth]{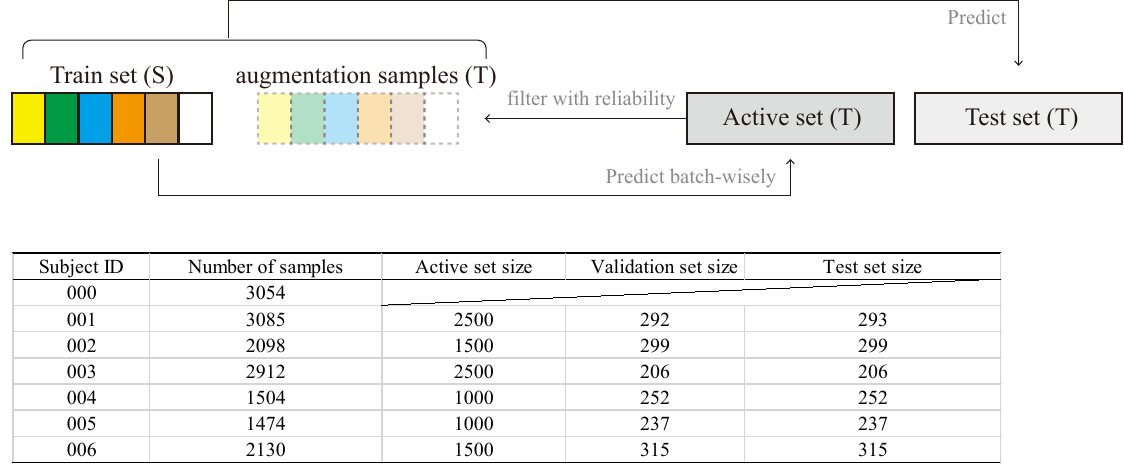}
    \caption{\textbf{The study pipeline of unsupervised cross-user adaptation.} 'S' stands for the source domain, and 'T' is for target domain.}
    \label{study pipeline}
\end{figure*}

To investigate the two unsupervised cross-user adaptation methods, an online data augmentation study pipeline was designed in this study. As Fig. \ref{study pipeline} presents, a model trained with data from source domain ('S') consecutively predicted and absorbed the unlabelled data from target domain ('T') in a batch-wise manner, and the enlarged training dataset was finally used to build a new model to predict the test set in target domain. As an example to study the cross-user data distribution drift, we chose the subject 000 as the subject in source domain, and the other subjects (001-006) were assumed as the target users. Based on the training data from subject 000, the augmentation scheme was independently applied for each subject, from 001-006. The data from subject 001-006 were randomly partitioned into active set (the unlabeled data in the target domain used for training data augmentation), validation set, and test set, and the scale of dataset partition is listed in Fig. \ref{study pipeline}. To statistically investigate the effectiveness of cross-user adaptation methods, parallel experiments with 30 times of random dataset partition were carried out, where the classification accuracy was considered the main metric.

\subsection{Unsupervised Cross-user Adaptation with Conformal Prediction with \\ Shrunken Centroids (CPSC)}
Conformal prediction (CP) is a computational framework effective in prediction reliability quantification \cite{CPSC_tutoria_2,IETBRAIN,CPSC_tutorial}. Instead of directly modeling the conditional probability, CP models the conformity of a particular feature-label combination with the training data based on a nonconformity measurement. Based on the nonconformity measurement, for a specific sample, CP gives a p-value indicating the conformity of each possible label. Then, two reliability metrics can be calculated based on the p-values: 1) credibility: the largest p-value, which reflects the risk of choosing the best label; 2) confidence: 1 minus the second largest p-value, which reflects the risk of excluding the other potential labels.

CP has been recently shown to improve the electronic-nose-based pattern recognition tasks with the selection of more reliable predictions on the unlabeled data \cite{CP_application,liu2021boost,IETBRAIN,lungcancer}. For example, with simulated data distribution drift from labeled training data to unlabeled active, validation and test data, the data augmentation strategy with the CP based on k-nearest neighbours (CPKNN) significantly improve the model accuracy on classifying different types of alternative herbal medicines \cite{liu2021boost}. 

Furthermore, we have proposed the novel framework called conformal prediction with shrunken centroids (CPSC) to optimize the reliability quantification in high-dimensional sensor data with higher augmentation effectiveness and significantly reduced the computational time \cite{liu2021cpsc}. The major characteristic of CPSC is that the model attenuates the noisy dimensions of features in the high-dimension space and therefore reduces the actual computational dimensionality \cite{sc}. Furthermore, different from the conventional CPKNN, CPSC only computes the Euclidean distance between a new sample and the class centroids, which significantly reduces the computational time needed \cite{liu2021cpsc}. These two characteristics are very helpful in dealing with the sEMG taste recognition problem, as the features are of high-dimensionality and there are thousands of samples, which may suffer from the curse of dimensionality and lengthy computational time with the conventional CPKNN. Therefore, in this study, we adopted the CPSC and data augmentation strategy to address the cross-user data drift by leveraging the unlabeled data in the target domain. 

With CPSC, to improve the taste recognition classifier performance across users, we regarded the training data from a user as the labeled training data (source domain) and partitioned the data from another user (target domain) into active set, validation set and test set. The active set will be partitioned evenly into four batches and the CPSC is used to predict the labels of the active samples in a batch-wise manner. The predictions with their predicted labels which satisfy the following two requirements are accepted as the augmentation data to be added into the current training data:

\begin{itemize}
\item The prediction credibility is larger than a threshold $\epsilon$
\item The largest p-value is larger than 3 times the second largest p-value;
\end{itemize}

These two requirements ensure that the selected augmented samples are with high credibility values and confidence values which are not very low. Then, the augmented training dataset will be used to predict the next batch of active samples and absorb new augmentation data. Benefited from the unlabeled data from the target domain, the final augmented training set will be used to predict the validation samples and test samples to evaluate the model accuracy.

It worths noting that during the entire process, the data on the target domain (the target user) were used in an unsupervised manner since the ground-truth labels on the target domain were constantly unknown to the models throughout the reliability-based augmentation process.

\subsection{Unsupervised Cross-user Adaptation with Domain Regularized Component Analysis}
One of the major challenges of sEMG-based taste recognition is the cross-user data drift. Since the data collected from different users can be highly dependent on individual user's skin characteristics, directly applying a model trained on one user's data collected in the laboratory may lead to inferior performance on the real-world test samples collected from another user under a different environment. To compensate for the domain drift, we leveraged the domain regularized component analysis (DRCA) \cite{zhang2017anti} to find a common projection hyper-plane from the source domain (the labeled training data from one user) and the target domain (the unlabeled active/validation/test data from another user) with convex optimization and the linear projection, which minimizes the difference of the data distributions from the two domains. If we denote the samples on the source domain as $x_i^S \in \rm I\!R^{D}, i=1,2,...,N^S$  and the samples on the target domain as $x_i^T \in \rm I\!R^{D}, i=1,2,...,N^T$, where $D$ denotes the original feature dimension, the summary statistics of the distribution of the data can be calculated: 

1) the mean vectors for the source-domain data $\mu^S = \Sigma_{i=1}^{N^S}x_i^S \in \rm I\!R^{D}$, for the target-domain data $\mu^T = \Sigma_{i=1}^{N^T}x_i^T \in \rm I\!R^{D}$ and for all data: $\mu = \frac{N^S \times \mu^S + N^T \times \mu^T}{N^S+N^T}$;

2) the within domain scatter for the source-domain data $S_w^S \in \rm I\!R^{D \times D}$ and for the target-domain data $S_w^T \in \rm I\!R^{D \times D}$ are calculated by: 
\begin{equation}
\begin{aligned}
S_w^S = \Sigma_{i=1}^{N^S}(x_i^S-\mu^S)(x_i^S-\mu^S)^T  \\
S_w^T= \Sigma_{i=1}^{N^T}(x_i^T-\mu^T)(x_i^T-\mu^T)^T 
\end{aligned}
\end{equation}

3) the between domain scatter $S_b \in \rm I\!R^{D \times D}$ is calculated by: 
\begin{equation}
S_b = N^S \times (\mu^S - \mu)(\mu^S - \mu)^T + N^T \times (\mu^T - \mu)(\mu^T - \mu)^T  \in \rm I\!R^{D \times D}
\end{equation}

Qualitatively, the mean vectors describe the overall distribution in terms of the centers of the data from source domain and the data from target domain. The within domain scatter $S_w^S$ and $S_w^T$ describes the spread of the data within each domain, and the between domain scatter $S_b$ described the spread between the source domain and the target domain with the mean of all samples as the reference point. 

With these summary statistics, the goal of the DRCA is to find a projection matrix $P \in \rm I\!R^{D \times \Tilde{D}}, \Tilde{D} < D$ to reduce the between domain scatter while maintaining the within domain scatter of the data on the projected hyper-plane. The rationales of DRCA are: 1) the minimized domain difference may enable the finding of the features shared by both domains that are robust to domain shift and effective in the classification of different taste sensations; 2) the maximized data spread from each domain enables the information and variance within the data to be kept for further classification modeling. 

Suppose the sample $x_i$ is projected onto $\Tilde{x_i}$:  $\Tilde{x_i} = P^T x_i \in \rm I\!R^{\Tilde{D}} $. Then, the same three types of summary statistics can be calculated on the projected data. For example, the within domain scatter on the projection hyper-plane are represented as: $\Tilde{S_w^S} = P^TS_w^SP \in \rm I\!R^{\Tilde{D} \times \Tilde{D}}$ and $\Tilde{S_w^T} = P^TS_w^TP \in \rm I\!R^{\Tilde{D} \times \Tilde{D}}$. The between domain scatter on the projection hyper-plane is represented as: $\Tilde{S_b} = P^TS_bP \in \rm I\!R^{\Tilde{D} \times \Tilde{D}}$. Therefore, to minimize the domain difference while maintaining the data spread from each domain of data on the projection hyper-plane, we design our optimization problem as the follows:
\begin{equation}
    \mathrm{max_P} \frac{tr(\Tilde{S_w^S} + \alpha \Tilde{S_w^S})}{tr(\Tilde{S_b})}
\end{equation}

The $\alpha$ is a hyperparameter to weigh the data from source domain and the data from target domain, which needs to be tuned for the best model performance. By expanding the expressions of the summary statistics on the projection hyper-plane, the problem can be reformulated as: 
\begin{equation}
    \mathrm{max_P} \frac{tr(P^TS_w^SP + \alpha P^TS_w^TP)}{tr(P^TS_bP)}
\end{equation}

The convex optimization problem in the fractional format can be then reformulated as a constrained optimization problem:

\begin{equation}
\begin{aligned}
    \mathrm{max_P} tr(P^TS_w^SP + \alpha P^TS_w^TP) \\
    s.t. tr(P^TS_bP)=\lambda
\end{aligned}
\end{equation}

Upon applying Lagrangian multiplier for this convex optimization problem, the Lagrangian can be represented as:
\begin{equation}
    L(P, \theta) = tr(P^T S_w^S P + \alpha P^T S_w^T P) - \theta (tr(P^T S_b P) - \lambda)
\end{equation}

By taking the derivative of the Lagrangian with respect to $P$ and setting it to zero, the problem can be solved as an eigenvalue decomposition problem:
\begin{equation}
    \begin{aligned}
        \frac{\partial L(P, \theta)}{\partial P} = 2 (S_w^S + \alpha S_w^T) -  2 \theta S_b = 0\\
        S_b^{-1}(S_w^S + \alpha S_w^T) = \theta P
    \end{aligned}
\end{equation}

The $P$ denotes the eigenvector matrix and by ranking the eigenvectors based on the eigenvalues, we can extract the $\Tilde{D}$ eigenvectors associated with the $\Tilde{D}$ largest eigenvalues and perform projection onto these eigenvectors. Then, the data matrix after projection will be changed from $N^S \times D$ and $N^T \times D$ to $N^S \times \Tilde{D}$ and $N^T \times \Tilde{D}$.

\subsection{Statistical Tests}
To test the robustness of the results and whether there is statistically significant accuracy improvement brought by CPSC and/or DRCA, we performed the Wilcoxon signed-rank tests based on the classification accuracy records from the 30 parallel experiments. The paired t-test was not used because the Shapiro-Wilk test rejected the normal distribution assumption on some results and the Wilcoxon signed-rank tests do not rely on the data normality assumption. Significance level was set as 0.05.

\section{Results}

\begin{figure*}

    \centering
    \includegraphics[width=\linewidth]{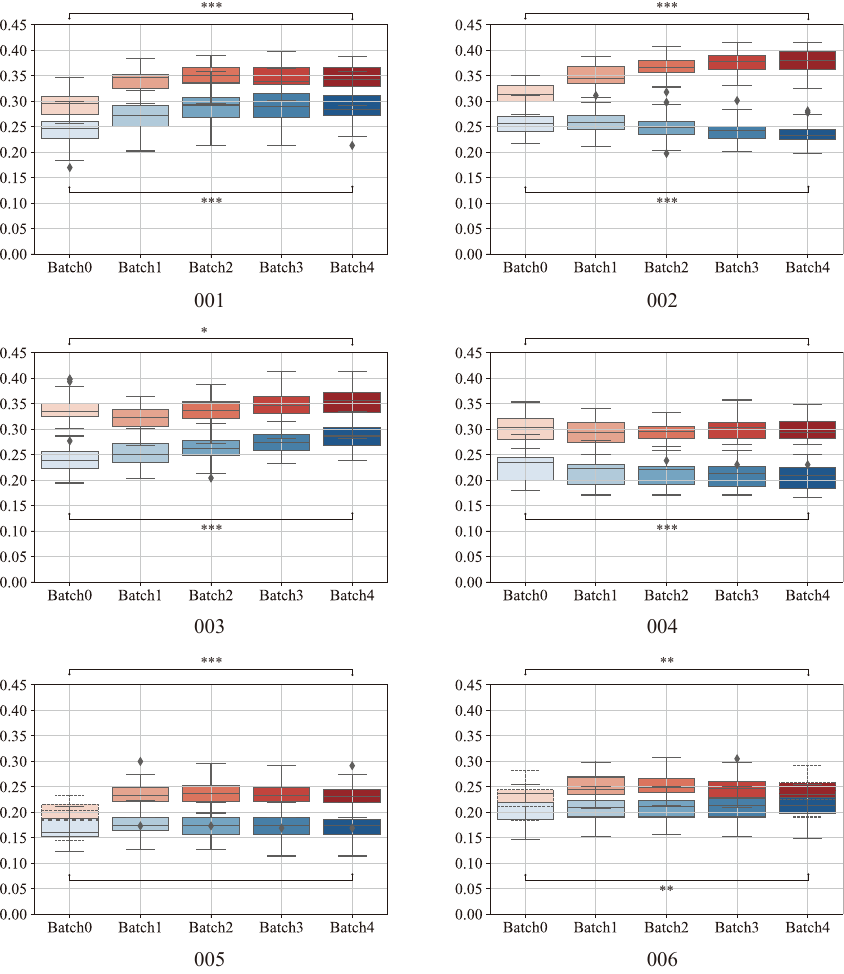}
    \caption{\textbf{The prediction accuracy of different models and the statically significant difference with Wilcoxon signed-rank tests.} 001-006 denotes the experiment results for the six subjects respectively, where the red boxes are the results gained after we performed DRCA to project data onto the lower-dimensional hyper-plane, while the blue ones are the results in the original feature space. The asteroid denotes the statistically significant difference level: '***' is when p-value $\textless$ 0.001, '**' is when p-value $\textless$ 0.01, while '*' is when p-value $\textless$ 0.05.}
    \label{fig_boxplot}
\end{figure*}

\begin{figure*}

    \centering
    \includegraphics[width=\linewidth]{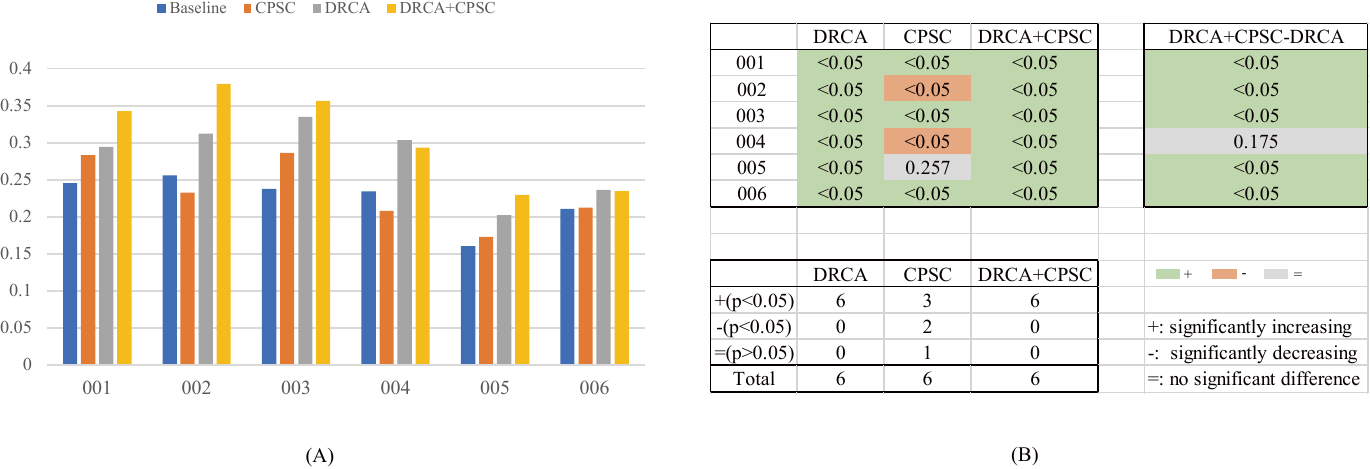}
    \caption{\textbf{The summarized result of median accuracy under six different tasks.} A) 001-006 denotes the task when each one of the six subjects was regarded as the user (target domain). The lines present the median accuracy with different models, over 30 parallel experiments with random partitions. B) The p-values given by the Wilcoxon signed-rank test, where the baselines (Batch 0 without DRCA) are compared with the results of 'DRCA'(Batch 0 with DRCA), 'CPSC'(Batch 4 in the original feature space without DRCA), 'DRCA+CPSC'(Batch 4 with DRCA), and their effects played on six subjects are also summarized in terms of the counts of subjects with significantly increasing accuracy/significantly decreasing accuracy/no significant difference. Furthermore, the comparisons before/after augmentation with CPSC on lower-dimensional hyper-plane are also investigated, denoted as 'DRCA+CPSC-DRCA'.}
    \label{fig_radar}
\end{figure*}

To better present the results in this study, some abbreviations of the terms are used in the paragraphs below, which are shown in Tabel. \ref{abbreviation}. For each subject in 001-006, the baseline model was the same to be compared against CPSC and DRCA: they were trained on the data from subject 000 in original feature space. Then we examined the models' performance variations with DRCA, CPSC, and DRCA+CPSC in the adaptation process from a dynamic perspective.\\
\indent The classification accuracy variation of the models with each batch is illustrated in Fig. \ref{fig_boxplot}. The red boxes and blue boxes respectively exhibit the results in the accuracy on the lower-dimensional hyper-plane (found by  DRCA) and the ones in the original feature space (without DRCA). To summarize the result, the median accuracy of the models over 30 parallel experiments are presented in Fig. \ref{fig_radar}, and the statistically significant changes of each subject (task) are counted.

\begin{table*}[]
\centering
\caption{The abbreviation of the terms used in result session.}
\scalebox{1}{

\begin{tabular}{ccl}
\cline{1-2}
Abbreviation & Meaning                                                                                      &  \\ \cline{1-2}

DRCA  & models with the data projected on the lower-dimensional hyper-plane                                                    &  \\
CPSC  & models with reliability-based data augmentation by CPSC on the original feature space                                      &  \\
\multicolumn{1}{l}{DRCA+CPSC} & \multicolumn{1}{l}{models with the augmentation process by CPSC on the lower-dimensional hyper-plane} &  \\ \cline{1-2}
\end{tabular}}
\label{abbreviation}
\end{table*}

According to Fig. \ref{fig_boxplot} and \ref{fig_radar}, on the six tasks on the original feature space (without DRCA), CPSC manifested statistically significant improvement on three subjects (p $\textless$ 0.05), statistically significant decrease on two subjects (p $\textless$ 0.05), and non-decreasing effect on one subject. On the lower-dimensional hyper-plane (with DRCA), CPSC generally exhibits a statistically significant accuracy improvement (p $\textless$ 0.05 on subjects 001-003,005-006; p=0.175 on subject 004). As for the dynamics of the accuracy variation, the augmentation process on lower-dimensional hyper-plane presents a similar pattern across subjects: as the number of batches of unlabeled target domain data increases, the classification gradually improves.   
When compared with the baseline models, the models with DRCA (projection data on lower-dimensional hyper-plane) exhibit statistically significant improvement in classification accuracy on every batch and on every subject (p $\textless$ 0.05). The  improvement on the median accuracy varied among subjects, from 0.025 (subject 006) to 0.097 (subject 003).

With DRCA+CPSC, the models achieve remarkable improvements when compared with the baseline on all subjects, with statistically significance (p $\textless$ 0.05). Respectively, the median classification accuracy was improved from 0.26 to 0.34 (subject 001), 0.26 to 0.38 (subject 002), 0.24 to 0.36 (subject 003), 0.23 to 0.29 (subject 004), 0.16 to 0.23 (subject 005), 0.21 to 0.23 (subject 006).

\section{Discussion}
In this study, we have leveraged the domain regularized component analysis (DRCA) and conformal prediction with shrunken centroids (CPSC) to address the cross-user data distribution drift problem in taste sensation recognition with surface electromyography (sEMG). Based on the sEMG data from six volunteers, we set the data from one volunteer as the source domain, and the data from the other six volunteers (subjects) as the target domain data respectively. Upon partitioning the target domain data into the active data, validation data and test data, we leveraged the unlabeled active data to perform the DRCA to find the hyper-plane with linear projection where the projections from the source domain and from the target domain data are closer while the data variance is maintained. After projecting the data onto the hyper-plane, we applied the CPSC as the classifier and adopted the reliability-based data augmentation with the unlabeled active data according to the protocol previously published \cite{liu2021cpsc,liu2021boost}. The results show that the DRCA significantly improve the cross-user taste sensation recognition accuracy while the improvement given by CPSC is not guaranteed. On three of the six volunteers regarded as the target domain subjects, the CPSC reaches non-decreasing accuracy while on two of the six volunteers, the CPSC leads to an significant decrease in accuracy. However, after using DRCA, the CPSC can always lead to non-decreasing accuracy (significant accuracy increase in 5/6 volunteers, when compared with the models with DRCA only). Therefore, the DRCA and CPSC can be used in combination to solve the cross-user data drift problem in sEMG taste sensation recognition problems in an unsupervised manner (i.e. without using the labels of the data from the target domain), for better classification performances in real-world applications.

The two major strategies we proposed to address the cross-user data drift issue are DRCA and CPSC, which worth further discussion. Firstly, the DRCA applies a linear projection onto the original high-dimensional data to project the data onto a lower-dimensional hyper-plane to make the projected data less separated in terms of between-domain scatter, while the within-domain scatter is kept for more information in the data to enable better classification performance. The rationale of this approach is to find a shared subspace that shares the commonality across the two domains for sake of compensating the difference and maintaining the similarities. This approach relies on the assumption that the subspace contains the taste sensation classification information which are invariant across different users and the domain drift between users are relatively trivial for taste sensation classification. The effectiveness shown in the statistically significant accuracy improvement in our study has proves that the assumption is satisfied and it is possible to find a cross-user invariant subspace to compensate the data drifts for better cross-user classification performance. 

Secondly, the CPSC strategy seeks to improve the cross-user classification performance in a semi-supervised manner: the CPSC strategy builds a model on the training data from the source domain, predicts the labels of the target domain active data, quantifies the reliability of the predictions in terms of credibility and confidence, filters the predictions with high credibility and then augments the training data with the selected samples and their predicted labels. When compared with the DRCA, the CPSC addresses the cross-user drift in a less direct manner. This reliability-based unlabeled data augmentation process assumes that the incorporation of unlabeled target domain data enables the models to gradually adapt to the data distribution on the target domain. However, as the reliability constraints for the prediction filtering are based on the training data under supervision, the "inertia" of this approach may be higher: the models may rely too heavily on the data distribution of the source domain. As long as the data distribution drifts become more significant, this approach may not be effective enough to improve the cross-user classification performance. This can be shown in the comparison between this study and our two previous studies \cite{liu2021cpsc,liu2021boost}, the data distribution drifts in this study may be more evident as the data are directly collected from different users at different times, while the data in the previous studies are either collected from the device at approximately the same time periods (more homogeneous) or simulated with Gaussian noises addition to the homogeneous data. Although the CPSC performs well in two of our previous studies, the more heterogeneous data distribution in this study renders the reliability-based unlabeled data augmentation strategy fail to guarantee the significantly increasing model accuracy in this study. 

To sum up, the DRCA strategy generally copes with the cross-user data distribution drift more directly than the CPSC strategy. Upon using the DRCA to project the data onto the lower-dimensional hyper-plane, the addition of CPSC can generally lead to an significantly increase cross-user classification accuracy. According to this study, it is suggested the users can combine the DRCA and CPSC strategies together to ensure higher cross-user classification accuracy according to the results in this study.


In this study, we situate our algorithms in the application of sEMG-based taste sensation recognition. The taste sensation based on sEMG can enable better human feeding feedback without any verbal communication and enables those who cannot verbally express themselves to convey their taste sensation information to the outside world, such as those suffering acute traumatic brain injury in the ICU who cannot verbally express their taste feelings. Furthermore, for the artificial taste stimulation system \cite{Wang2,Wang3,Wang4}, this system enables the feedback quality control for the system fine-tuning to provide the users with more bona fide taste sensation experience. In the sEMG-based taste sensation recognition task, cross-user data distribution drift is very common due to the individual differences in electrophysiological characteristics, the different environmental temperature and humidity, and the different situations of the sensors themselves. Furthermore, although the standard solutions with fixed concentrations can be used for volunteers in laboratory tests, when used in real-world applications, the system may also suffer from the non-standard taste stimulation. Theses different factors can contribute to significant decrease in taste sensation recognition accuracy in real-world applications. Therefore, the strategies to compensate the domain drift is very important for the utilization of this sEMG-based taste sensation recognition system outside the laboratory environment.

Beyond the application in sEMG-based taste sensation recognition, the DRCA combined with CPSC can be applied in broader range of sensor-based tasks which involves the domain adaptation and data distribution drifts. For example, this strategy proposed in this study can be applied to the volatile organic compound pattern recognition with electronic nose, where the sensor drifts caused by environment temperature, humidity and sensor aging and sensor poisoning can lead to significantly inferior classification rate  \cite{IETBRAIN,lungcancer,FeatureEngineering2019}. Furthermore, this novel domain adaptation strategy can also be applied to more biomedical applications other than the sensor-based fields, such as the COVID-19 outcome prediction tasks considering the data distribution drifts across different virus variants \cite{covid}.

Although this study has shown the effective domain adaptation with DRCA and CPSC on sEMG-based taste sensation recognition task, there are several limitations that need to be mentioned and addressed in the future studies. Firstly, in this study, the CPSC strategy does not robustly result in improved cross-user classification accuracy, which indicates that the CPSC may not be able to focus on the features invariant to domain drift. In the future, more complicated and flexible conformal prediction models, such as the conformal prediction based on one-dimensional convolutional neural network (CP-1DCNN), conformal prediction based on recurrent neural network with long short time memory (CP-LSTM), can be developed and tested to optimize the cross-user classification accuracy for this specific task since they have shown high effectiveness in the sensor-signal-based classification tasks \cite{liu2021classifying}. 

Secondly, even though the DRCA reaches significantly higher accuracy than the baselines and CPSC strategy, the overall classification accuracy is still not very high. Therefore, further feature engineering may be needed in the future to improve the feature extraction process to optimize the classification accuracy \cite{liu2021classifying}. More features such as the spectral densities or discrete-wavelet-transform-based features can be tested on this task. 

Thirdly, although this study models the data distribution drift with the data from different volunteers, this protocol only models the domain drift caused by different subjects. In the future, more experiments involving different volunteers and different sEMG sensors are warranted to test the effectiveness of the DRCA and CPSC strategies when more factors of data drifts are taken into consideration. 

Furthermore, in this study, we applied the DRCA as a simple linear projection approach to address the cross-user data drift problem. In the future, more flexible and non-linear approaches such as the generative adversarial network (GAN), the transfer learning technology can be tested on this task for optimized cross-user classification. 

Additionally, in this study, to get larger quantities of data, we used the sliding window to extract frames from the sEMG experiment signals for each volunteer as the samples. As a sacrifice, this approach generally leads the samples to violate the assumption that the data are sampled independently and identically from the distribution. The potential correlation within the data from the same volunteer may leads the model to be less generalizable. In the future, with more experimental data, other sampling methods can be used to extract samples from the sEMG signals.

\section{Conclusion}
In this study, we proposed two strategies to address the issue of cross-user data drift from one user (source domain) to another user (target domain) in the sEMG-based taste sensation recognition applications: domain regularized component analysis (DRCA) and conformal prediction with shrunken centroids (CPSC). DRCA finds a projection which maps the data from two domains into a shared lower-dimensional hyper-plane, and CPSC provides reliability information when giving predictions which enables reliability-based unlabeled data augmentation. When the strategies were applied independently, the DRCA significantly improved the cross-user classification accuracy while the CPSC did not significantly improve the accuracy when compared with that achieved by the baseline models trained on the source domain training data. When the two strategies were combined, the classification accuracy was significantly improved when compared with those based on both the baseline models and the models with DRCA only. Therefore, we suggest that the users can use both strategies to solve the cross-user data distribution drift issue for a more accurate application of taste sensation recognition. Further studies are warranted to extract better features to improve the overall cross-user classification accuracy for more robust and reliable real-world applications.

\section*{Acknowledgment}
The authors would like to thank all participants who contributed to our project; special thanks to Qing Ai, Xinyu Li, Lizi Jiang, Hongyin Wei, Bixuan Zhang, and Zhang Ming for their kind help during data acquisition.

\ifCLASSOPTIONcaptionsoff
  \newpage
\fi

\bibliographystyle{IEEEtran} 
\bibliography{cited}

\end{document}